\documentclass[aps,nofootinbib]{revtex4-1}
\usepackage{graphicx,subfigure}
\usepackage{amsmath,amssymb}
\usepackage{color}

\usepackage{graphicx,subfigure}

\def\beq{\begin{equation}}
\def\eeq{\end{equation}}
\def\beqr{\begin{eqnarray}}
\def\eeqr{\end{eqnarray}}
\def\bdpm{\begin{displaymath}}
\def\edpm{\end{displaymath}}
\def\half{\frac{1}{2}}

\begin{document}

\title{
Singlet fermionic dark matter with Veltman conditions
}

\author{Yeong Gyun Kim}
\email{ygkim@gnue.ac.kr}

\affiliation{ 
Department of Science Education, 
Gwangju National University of Education, Gwangju 61204, Korea
}

\author{Kang Young Lee}
\email{kylee.phys@gnu.ac.kr}

\affiliation{ 
Department of Physics Education \&
Research Institute of Natural Science,
Gyeongsang National University, Jinju 52828, Korea
}

\author{Soo-hyeon Nam}
\email{glvnsh@gmail.com}

\affiliation{ 
Department of Physics, 
Korea University, Seoul 02841, Korea
}

\date{\today}

\begin{abstract}

We reexamine a renormalizable model of a fermionic dark matter with a gauge singlet Dirac fermion and a real singlet scalar 
which can ameliorate the scalar mass hierarchy problem of the Standard Model (SM).
Our model setup is the minimal extension of the SM for which a realistic dark matter (DM) candidate is provided and 
the cancellation of one-loop quadratic divergence to the scalar masses can be achieved 
by the Veltman condition (VC) simultaneously.
This model extension, although renormalizable, can be considered as an effective low-energy theory
valid up to cut-off energies about 10 TeV.
We calculate the one-loop quadratic divergence contributions of the new scalar and fermionic DM singlets,  
and constrain the model parameters using the VC and the perturbative unitarity conditions. 
Taking into account the invisible Higgs decay measurement, 
we show the allowed region of new physics parameters satisfying the recent measurement of relic abundance.
With the obtained parameter set, 
we predict the elastic scattering cross section of the new singlet fermion 
into target nuclei for a direct detection of the dark matter. 
We also perform the full analysis with arbitrary set of parameters without the VC as a comparison,
and discuss the implication of the constraints by the VC in detail.

\end{abstract}

%\pacs{PACS numbers:12.60.Fr,12.60.Cn,14.80.Cp}
\maketitle

%%%%%%%%%%%%%%%%%%%%%%%%%%%%%%%%%%%%%%%%%%%%%%%%%%%%%%%%%%%%%%%%%%%%%%%%
\section{Introduction}
%%%%%%%%%%%%%%%%%%%%%%%%%%%%%%%%%%%%%%%%%%%%%%%%%%%%%%%%%%%%%%%%%%%%%%%%

Recent discovery of a Higgs-like boson at the LHC \cite{ATLAS12,CMS12}
completes the standard model (SM) as a renormalizable theory,
and its observed mass $M_h = 125.9 \pm 0.4$ GeV validates
the perturbativity of weak interactions in the model \cite{Lee77}.
In the SM, a scalar mass receives additive quantum corrections
proportional to the momentum cut-off squared, $\Lambda^2$ at one-loop level.
This quadratically divergent one-loop correction to the Higgs mass was first calculated by Veltman \cite{Veltman81},
and given by 
\beq \label{eq:1loophiggsmass}
\delta M_h^2
= \frac{3\Lambda^2}{16\pi^2v_h^2}
\left(M_h^2 + 2M_W^2 + M_Z^2 -4m_t^2\right),
\eeq
where $v_h$ is the vacuum expectation value (VEV) of the Higgs field
and only the dominant contribution is shown in the scale $\Lambda$.
If $\Lambda$ is very large,
the Higgs mass suffers from the quadratic divergence of this correction,
which addresses the naturalness issue of the theory
resulting in the fine-tuning problem \cite{Susskind79}.
Since the cut-off scale $\Lambda$ is considered to be UV boundary of the SM \cite{Georgi74},
it would be more natural that the cancellation of the quadratic divergences could be achieved
by means of a symmetry principle at higher scales
rather than by fine-tuning of mathematically supplemented counterterms.
There have been various extensions of the SM
to resolve the fine-tuning (or the hierarchy) problem of the Higgs mass
by canceling the quadratic divergences
such as the supersymmetry (SUSY) and the little Higgs model (LHM).
However no compelling sign of evidence for new physics (NP)
beyond the SM has been found at the LHC.

Veltman has suggested
the cancellation of the quadratic divergence of the Eq. (\ref{eq:1loophiggsmass}) by itself
leading to the prediction of the Higgs mass, 
which is called the Veltman condition (VC).
It might imply the existence of some hidden symmetry though the underlying theory is unknown.
The simple VC predicted $M_h \sim 320$ GeV
with the known particle masses at present
,and does not hold in the SM with the present measured value $M_h \sim 125$ GeV.
In order to impose the VC to the Higgs mass,
required are some new degrees of freedom
which interact with the Higgs boson to contribute to the Higgs self-energy diagram \cite{Gunion08}.
Simple extensions of the SM resolving the fine-tuning problem by applying the VC
have been studied widely in many literatures 
\cite{Grzadkowski09,Pivovarov08, Masina13, Karahan14,Demir15}. 

Besides the fine-tuning issue,
there are other motivations to search for the NP beyond the SM.
One of the most important motivation is
the existence of the distinct observational evidence for dark matter (DM).
Thus many viable DM candidates have been suggested in various NP models at present.
In order to include DM in the SM framework,
we should assume the existence of additional degrees of freedom.
Since any SM particle cannot be DM,
DM itself should be a new degree of freedom,
and new particles or new interactions are also required to connect the DM and the SM sector.
In the Higgs portal model for the DM,
only the Higgs quadratic term takes part in interacting with the new degrees of freedom.
Thus the Higgs portal model can provide both of the DM
and the interaction of new degrees of freedom with the Higgs boson to satisfy the VC for the Higgs mass.
Recently there has been some efforts to investigate a possibility
to explain the fine-tuning of the Higgs mass through the VC 
in the Higgs portal model for the dark matter.
One of the simplest extension to satisfy the VC is
to introduce only one new scalar to cancel the dominant top contribution
in Eq. (\ref{eq:1loophiggsmass}) \cite{Karahan14,Demir15}.
New scalar degrees of freedom for ameliorating the fine-tuning problem
could be DM candidates in itself interacting with the SM sector through Higgs portal
with a discrete $Z_2$ symmetry \cite{Drozd12}.
However, a new quadratic divergence to the new scalar DM mass
arises in this minimal model and we have to resolve it as well.
For instance, we may introduce additional vector-like fermions
to cancel the new quadratic divergence
and to protect the masses of the Higgs boson and the new scalar DM simultaneously
\cite{Chakraborty13}.

Alternatively we consider the Higgs portal model
with the singlet fermionic dark matter as a more general scenario in this paper
since this model was first introduced by our earlier studies \cite{Lee07}.
In this model, we have a Dirac fermion and a real scalar,
which are both the SM gauge singlets.
If we assign no $Z_2$ symmetry, the singlet scalar cannot be a DM
since it would decay into the SM sector through mixing.
Thus the DM is the singlet fermion in this model.
We impose the VC on the quadratic divergences of both scalar masses
of the SM-like Higgs boson and the singlet scalar in our model.
Under the VC, we still find the allowed parameter space
where the singlet fermion satisfies the observed relic density constraints
to be the DM candidates.
It implies that our model of the DM provides the improved naturalness
at least at one-loop level.
Since the VC is a very strong condition to fix the Higgs-scalar coupling $\lambda_2$ to a specific value,
only the limited parameter sets are allowed.
Also explored is the direct detections of the singlet fermion
through the DM-nucleon scattering under the VC
and the prospect of observation of the DM in the future is discussed in this paper.
All the analyses are performed with the up-to-date cosmological data 
without applying the VC as well
in order to clearly see the implication of the VC constraints and also for the general DM study 
as a reference.
This model might provide richer phenomenology especially in the studies
of the electroweak phase transition
responsible for the baryon asymmetry in the early universe \cite{PMS}.

This paper is organized as follows.
In Sec. II, we briefly review our model and summarize the model parameters.
We discuss the VCs for the scalar masses in Sec. III
and constraints from the measurements of Higgs decays in Sec. IV.
Studied are phenomenologies for the dark matter
such as the relic density and the direct detection through DM-nucleon scattering in Sec. V.
Finally we conclude in Sec. VI.

%%%%%%%%%%%%%%%%%%%%%%%%%%%%%%%%%%%%%%%%%%%%%%%%%%%%%%%%%%%%%%%%%%%%%%%%
\section{model}
%%%%%%%%%%%%%%%%%%%%%%%%%%%%%%%%%%%%%%%%%%%%%%%%%%%%%%%%%%%%%%%%%%%%%%%%

We adopt a dark sector consisting of a real scalar field $S$ 
and a Dirac fermion field $\psi$ which are SM gauge singlets studied first in Ref. \cite{Lee07,Kim16}.  
The dark sector Lagrangian with the renormalizable interactions is then given by
\beq \label{eq:DM_Lagrangian}
\mathcal{L}_{DM} = \bar{\psi}\left(i\partial\!\!\!/ - M_{\psi_0}\right)\psi 
+ \half \left(\partial_{\mu} S\right)\left(\partial^{\mu} S\right)  - g_S \bar{\psi}\psi S - V_{S}(S,H), 
\eeq
where the Higgs portal potential is
\beq \label{eq:VS_potential}
V_{S}(S,H) = \half M_S^2 S^2 + \lambda_1 H^{\dagger} H S + \lambda_2 H^{\dagger} H S^2 
+ \frac{\lambda_3}{3!}S^3 + \frac{\lambda_4}{4!}S^4 .
\eeq
Note that the singlet fermionic DM field $\psi$ couples only to the singlet scalar $S$, and
the interactions of the singlet sector to the SM sector arise only through the Higgs portal $H^{\dagger}H$.  
After electroweak symmetry breaking, the neutral component of the SM Higgs and the singlet scalar 
develop nonzero VEVs,  $\langle H^0 \rangle=v_h/\sqrt{2}$ and $\langle S \rangle = v_s$, respectively.

Minimizing the full scalar potentials $V_S + V_{SM}$, where
\begin{eqnarray}
	V_{SM} = -\mu^2 H^{\dagger} H + \lambda_0 (H^{\dagger} H)^2,
\end{eqnarray}
the scalar mass parameters $M_S^2$ and $\mu^2$ are expressed in terms of the scalar VEVs as follows \cite{PMS}
\beqr
	M_S^2 &=& - \left(\frac{\lambda_1}{2 v_s} + \lambda_2 \right)v_h^2 - \left(\frac{\lambda_3}{2v_s} + \frac{\lambda_4}{6}\right) v_s^2 ,
	\nonumber \\[1pt]
 	\mu^2 &=& {\lambda}_0 v_h^2 + (\lambda_1 + \lambda_2 v_s)v_s .
\eeqr
The neutral scalar fields $h$ and $s$ defined by $H^0=(v_h+h)/\sqrt{2}$ and $S=v_s+s$ are mixed to yield the mass matrix given by
\begin{eqnarray}
\mu_{h}^2 &=& 2 {\lambda}_0 v_h^2 , \nonumber \\[1pt]
\mu_{s}^2 &=& - \frac{\lambda_1 v_h^2}{2 v_s} + \frac{\lambda_3}{2} v_s + \frac{\lambda_4}{3} v_s^2 , \nonumber \\[1pt]
\mu_{hs}^2 &=&  (\lambda_1 + 2 \lambda_2 v_s) v_h.
\end{eqnarray}
The corresponding scalar mass eigenstates $h_1$ and $h_2$ are admixtures of $h$ and $s$,
\beq
\left( \begin{array}{c} h_1 \\[1pt] h_2 \end{array} \right) =
\left( \begin{array}{cc} \cos \theta &\ \sin \theta \\[1pt]
  	- \sin \theta &\ \cos \theta \end{array} \right)
\left( \begin{array}{c} h \\[1pt] s \end{array} \right) ,
\eeq
where the mixing angle $\theta$ is given by
\begin{eqnarray}
\tan \theta = \frac{y}{1+\sqrt{1+y^2}},
\end{eqnarray}
with $ \ y \equiv  2 \mu_{hs}^2 /(\mu_h^2 - \mu_s^2)$. 
After the mass matrix is diagonalized, 
we obtain the physical masses of the two scalar bosons $h_{1,2}$ as follows:
\begin{eqnarray}
M^2_{1,2} = \frac{\mu_h^2+\mu_s^2}{2}
            \pm \frac{\mu_h^2-\mu_s^2}{2}\sqrt{1+y^2},
\end{eqnarray}
where the upper (lower) sign corresponds to $M_1(M_2)$.
We assume that $M_1$ corresponds to the observed SM-like Higgs boson mass in what follows.

The singlet fermion $\psi$ has mass $M_\psi = M_{\psi_0} + g_S v_s$ 
as an independent parameter of the model since $M_{\psi_0} $ is just a free model parameter, 
and the Yukawa coupling $g_S$ measures the interaction of $\psi$ 
with singlet component of the scalar particles. 
In total, we have eight independent model parameters relevant for DM phenomenology. 
The six model parameters $\lambda_0, \lambda_1, \lambda_2, \lambda_3, \lambda_4$ and $v_s$ 
determine the masses $M_{1,2}$, the mixing angle $\theta$, 
and self-couplings of the two physical scalars $h_{1,2}$. 
Given the fixed Higgs mass $M_1$, 
we constrain seven independent NP parameters taking into account various theoretical consideration
and experimental measurements in the next section.

%%%%%%%%%%%%%%%%%%%%%%%%%%%%%%%%%%%%%%%%%%%%%%%%%%%%%%%%%%%%%%%%%%%%%%%%
\section{Veltman condition and Unitarity}

%%%%%%%%%%%%%%%%%%%%%%%%%%%%%%%%%%%%%%%%%%%%%%%%%%
\begin{figure}[!hbt]
	\centering%
	\subfigure[ ]{\label{sfdm_scalar} %
		\includegraphics[width=5.4cm]{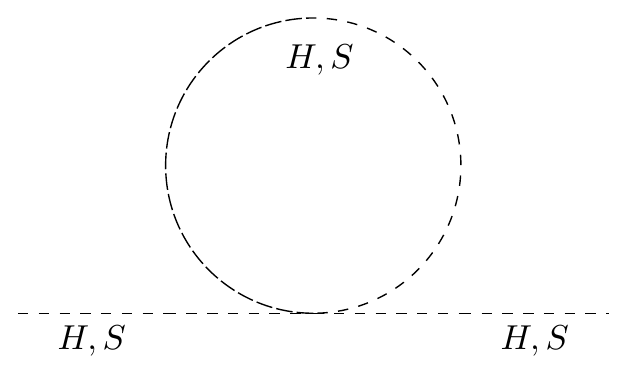}} \qquad  
	\subfigure[ ]{\label{sfdm_fermion} %
		\includegraphics[width=6cm]{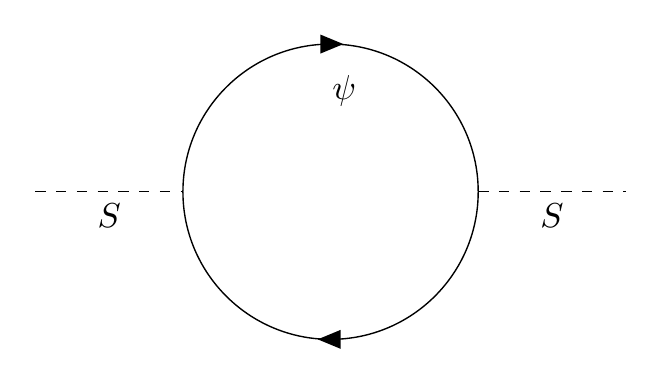}}	
	\caption{One-loop quadratic divergence contributions of the (a) scalars and (b) fermionic DM
		 to the Higgs and the new scalar masses from the hidden sector. } 
	\label{fig:oneloop}
\end{figure}
%%%%%%%%%%%%%%%%%%%%%%%%%%%%%%%%%%%%%%%%%%%%%%%%%%

We apply the VC to this model as done similarly in Refs.\cite{Chakraborty13,Karahan14,Demir15}, 
and constrain the above model parameters from the fine-tuning problem
 originating from the radiative corrections to the Higgs mass.
Cutting off the loop integral momenta at a scale $\Lambda$, and keeping only the dominant contributions in this scale, 
we obtain
\beq \label{eq:radhiggsmass}
M^2_h = (M^0_h)^2 + \frac{3\Lambda^2}{16\pi^2v_h^2}\left(M_h^2 + 2M_W^2 + M_Z^2 -4m_t^2 + \frac{\lambda_2}{3}v_h^2\right),
\eeq
where $M^0_h$ is the bare mass constrained in the unrenormalized Lagrangian,
and small mixing effect proportional to $\theta$ is neglected.
The last term proportional to $\lambda_2$ in Eq. (\ref{eq:radhiggsmass}) 
is obtained from the one-loop diagram in Fig. \ref{sfdm_scalar}. 
Following Veltman \cite{Veltman81},  note that by choosing the parameter $\lambda_2$ to be
\beq \label{eq:VC_Higgs}
\lambda_2 = 3(4m_t^2 - M_h^2 - 2M_W^2 - M_Z^2)/v_h^2 \sim 4.17,
\eeq
the quadratic divergences can be canceled and this could be a prediction for the parameter $\lambda_2$.
This result agrees well with the prior result of Ref. \cite{Demir15}
where the additional scalar mass is protected by a new hidden vector.
Also, the same result for $\lambda_2$ was obtained in the (multi-)scalar DM model studied 
in Ref. \cite{Chakraborty13}.	
On the other hand, in our model, a singlet fermionic DM stabilizes the mass of the scalar mediator.   
Similarly to the Higgs mass case, as for the scalar singlet mass, 
we obtain from Fig. \ref{fig:oneloop} that
\beq \label{eq:radscalarmass}
M^2_s = (M^0_s)^2 + \frac{\Lambda^2}{32\pi^2}\left(\lambda_4 + 2\lambda_2 - 8g_S^2 \right),
\eeq 
and the VC requires 
\beq \label{eq:VC_scalar}
\lambda_4 = 8g_S^2 - 2\lambda_2 .
\eeq
The conditions obtained above were given only at the one-loop level,
and cannot be the legitimate solution to the fine-tuning problem 
especially if the scale of NP is extremely large.
For scales not much larger than the electroweak scale, however,
one does not need very large cancellations.
If the Veltman solution is by chance satisfied,
the scale $\Lambda$ can be pushed at the two-loop level to a much higher value 
proportional to $\Lambda^2\log\Lambda$ for $\Lambda \sim 15$ TeV.
Nonetheless, including two-loop (or higher-loop) corrections will modify the above condition 
by further powers of $O(1/16\pi^2)$, 
so that a modification of the one-loop VC is very small at the level of a few percent \cite{Al-sarhi92}. 
Of course, one can set up the model parameters to apply the VC at the two-loop or higher level.   

The couplings $\lambda_{2,4}$ can grow significantly with increasing renormalization scale $\Lambda$,
and one can constraint those by applying the tree-level perturbative unitarity to scalar elastic scattering processes
for the zeroth partial wave amplitude \cite{Lee77}.
For a singlet scalar Higgs portal, the bounds on the couplings $\lambda_{2,4}$ are given by \cite{Cynolter04}
\beq \label{eq:unitarity}
|\lambda_2| \leq 4\pi  , \quad \lambda_4 \leq 8\pi .
\eeq
The value of $\lambda_2$ obtained from the VC in Eq.(\ref{eq:VC_Higgs}) lies obviously in the range of the above bound.
From the fact that the quartic coupling $\lambda_4$ should be positive, and from Eqs. (\ref{eq:VC_scalar}) and (\ref{eq:unitarity}) 
we obtain
\beq \label{eq:VC_coupling}
1.02 < g_S < 2.05 .
\eeq
Interestingly, the obtained size of allowed DM coupling $g_S$ is similar to 
that of the strong interaction coupling in the SM. 
Using the given conditions, we will perform the numerical analysis with the various sets 
of the following five NP parameters: 
$\sin\theta, M_2, M_\psi, g_S, v_s$.
In addition, we will also show the full analysis with arbitrary set of $\lambda_{2,4}$ without the VC 
as a comparison.

%%%%%%%%%%%%%%%%%%%%%%%%%%%%%%%%%%%%%%%%%%%%%%%%%%
\begin{figure}[!hbt]
	\centering%
	\subfigure[ ]{\label{fig:rge1} %
		\includegraphics[width=7.6cm]{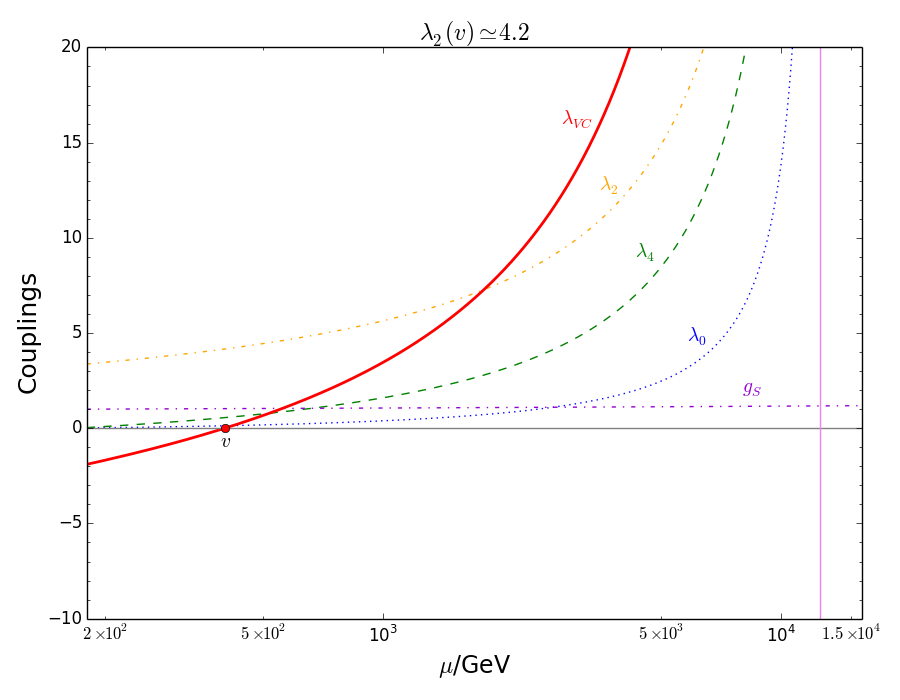}} \
	\subfigure[ ]{\label{fig:rge2} %
		\includegraphics[width=7.6cm]{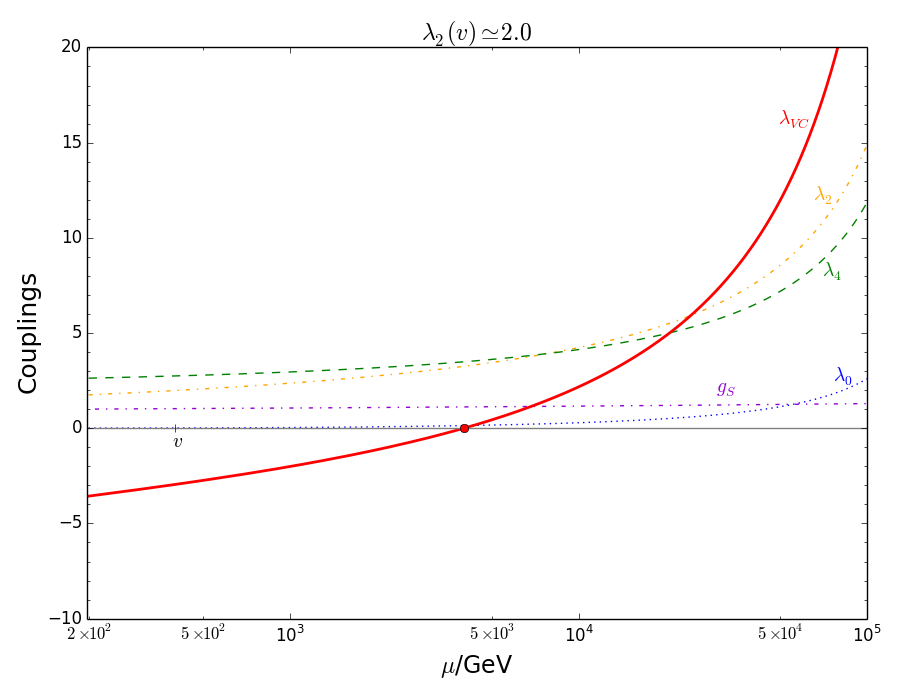}}  
	\caption{Running of the massless couplings and the first VC given in Eq. (\ref{eq:lambda_vc})
		for (a) $\lambda_2(v) = 4.17$ and (b) $\lambda_2(v) = 2.0$ 
		with the renormalization scale $\mu$.  
		The vertical(violet) line indicates where one of the couplings hits Landau pole.
		$g_S$ and $\lambda_4$ are chosen to satisfy the second VC obtained in Eq. (\ref{eq:VC_scalar}) 
		at (a) $\mu = v$ and (b) $\mu = 4$ TeV. } 
	\label{fig:rge}
\end{figure}
%%%%%%%%%%%%%%%%%%%%%%%%%%%%%%%%%%%%%%%%%%%%%%%%%%

Before we discuss the phenomenological aspects of this model in the next section,
we briefly check the renormalization group evolution (RGE) behavior of the model parameters. 	
The $\beta$-function of a coupling $\lambda_i$ at a scale $\mu$ in the RGE is defined as
 $\beta_{\lambda_i} = \partial \lambda_i/\partial \log\mu$.
For dimensionless couplings in the scalar potential (including the DM Yukawa coupling),
the one-loop $\beta$-functions are given by \cite{Baek12} 
\beqr
\beta_{\lambda_0}^{(1)} &=& \frac{1}{16\pi^2}\left[ 24\lambda_0^2 + 12\lambda_0\lambda_t^2 -6\lambda_t^4
	-3\lambda_0(3g_2^2 + g_1^2) +\frac{3}{8}\big(2g_2^4+(g_2^2+g_1^2)^2\big) + 2\lambda_2^2 \right],
\nonumber \\[1pt]
\beta_{\lambda_2}^{(1)} &=& \frac{\lambda_2}{16\pi^2}\left[ 12\lambda_0 
	-\frac{3}{2}\lambda_0(3g_2^2 + g_1^2) + 6\lambda_t^2 + 8\lambda_2 + \lambda_4 + 4g_S^2 \right],
\nonumber \\[1pt]
\beta_{\lambda_4}^{(1)} &=& \frac{1}{16\pi^2}\left[ 8\lambda_2^2 + \half \lambda_4^2 
	+ \frac{4}{3}\lambda_4 g_S^2 - 8g_S^4 \right], \qquad
\beta_{g_S}^{(1)} = \frac{5}{16\pi^2}g_S^3,
\eeqr	
where $g_1$ and $g_2$ are the SM U$(1)_Y$ and SU$(2)_L$ couplings, respectively,
and $\lambda_t$ is the top Yukawa coupling.
In terms of the running couplings, the quadratic radiative mass correction $\delta M_h^2$ in Eq. (\ref{eq:VC_Higgs})
can be rewritten at a scale $\mu$ as
\beq \label{eq:lambda_vc}
\delta M_h^2 (\mu) \equiv \frac{\Lambda^2}{16\pi^2}\lambda_{VC}(\mu)
	= \frac{\Lambda^2}{16\pi^2}\left(6\lambda_0(\mu) + \frac{9}{4}g_2^2(\mu) + \frac{3}{4}g_1^2(\mu) 
	- 6\lambda_t^2(\mu) + \lambda_2(\mu)\right).
\eeq
It is natural to determine the values of the physical observables 
at a scale $\mu = v \equiv \sqrt{v_h^2 + v_s^2}$ for a study of the DM phenomenology
since $v$ can be considered as the VEV of the radial component of a scalar field composed of $h$ and $s$.
Since we are interested in the DM fermion (and $h_2$) lighter than 1 TeV, 
we choose $v \sim 400$ GeV as a benchmark point.
In Fig. \ref{fig:rge}, we show the scale dependence on the RGE of the massless couplings 
together with the VC for two different set of parameter values:
Firstly, we consider a scenario in Fig. \ref{fig:rge1} 
that $\lambda_2(v) = 4.17$ and $\lambda_{VC}(v) \simeq 0$
as obtained in Eq. (\ref{eq:VC_Higgs}). 
The scalar couplings are drastically increased above 5 TeV and hit Landau pole near at 12.6 TeV. 
Therefore this DM model, although renormalizable, can be considered in this case
as an effective low-energy theory valid up to cut-off energies about 10 TeV. 
$\lambda_2(v) = 4.17$ is quite large value even if still perturbative.
Introducing $N$ number of singlets brings down the value to $4.17/N$
so that Landau pole can be shifted further to a much higher scale \cite{Chakraborty13}.
However we consider a simplest possible scenario ($N = 1$) as a reference. 
Secondly, we consider another scenario in Fig. \ref{fig:rge2} 
that $\lambda_2(v) = 2.0$. 
In this case, we found that $\lambda_{VC}$(4 TeV) $\simeq 0$ and 
one of the couplings hits Landau pole near at 300 TeV. 
At a scale $\mu = v$, the mass correction ratio becomes $\delta M_h^2/M_h^2 \sim 100$ 
which corresponds to fine-tuning at the level of $1\%$ following Ref. \cite{Kolda00}.
Satisfying the VC at the electroweak scale($\sim v$) is not mandatory.
The VC might possibly be satisfied at a higher NP scale or even at GUT scale.
However, it is considered to be more `natural' 
if the fine-tuning measure from the radiatively induced mass ratio is within the level of a few percent.
In this sense, the second scenario is also acceptable.
As for the other NP couplings,  
$g_S$ and $\lambda_4$ are chosen to satisfy the second VC obtained in Eq. (\ref{eq:VC_scalar}) 
at $\mu = v$ and $\mu = 4$ TeV in the first and second scenarios, respectively. 
In order for clear comparison with the prior results of Refs. \cite{Demir15, Chakraborty13} 
which have the same extra scalar (but different DM sector) resulting in the same prediction 
on $\lambda_2(v) = 4.17$, 
we will focus on the first scenario with a simplest possible setup of the DM sector for the numerical analysis.  
Nevertheless, the second scenario ($\lambda_2(v) = 2.0$) will not change our numerical results much.

The resolution of the fine-tuning issue suggests new weakly coupled physics 
at the cutoff-scale $\Lambda \sim$ 1 TeV in order to explain the currently measured Higgs mass. 
In our model, the first and the second scenarios push the cutoff to $\sim$ 10 TeV and 100 TeV, respectively.
Similarly, as an effective theory, the LHM addressing the fine-tuning problem 
also allow the cutoff of the theory to be raised up to $\sim$ 10 TeV,
beyond the scales probed by the current precision data. 
Such extensions of the SM (including SUSY) require new TeV scale particles 
in order to cancel the one-loop quadratic divergences.
However, the current electroweak precision measurements put significant constraints on those new particles, 
which indicate that there is no new particle up to $\sim$ 10 TeV
(unless one amends the models with extra discrete symmetries). 
This discrepancy creates a tension known as the little hierarchy problem \cite{Barbieri99}.
However, in our model, the Higgs portal scalar can alleviate the little hierarchy problem, 
and no new TeV scale particles are necessary in the electroweak sector.
Nevertheless, because of the non-perturbativity of the scalar quartic couplings 
at a few TeV scale in the first scenario ($\lambda_2(v) = 4.17$)  
one might want to invoke some NP in this case in order to reduce the quartic couplings at a higher scale. 
Even so, as far as such an NP extension resides only in the hidden sector,
LHC constraints on the NP extension shall be negligible for small $\sin\theta$ 
such as the case of $h_2$ and DM fermion phenomenology as we will discuss in the next chapters.
Besides all those, since we do not know either what kind of NP enters at a higher scale for UV completion of theory
or at which scale the VC is satisfied,  
we also perform the full analysis with arbitrary set of $\lambda_{2,4}$ without the VC as a comparison,
so that our numerical results can be applicable to various extensions of the fermionic DM models.

%%%%%%%%%%%%%%%%%%%%%%%%%%%%%%%%%%%%%%%%%%%%%%%%%%%%%%%%%%%%%%%%%%%%%%%%
\section{Collider Phenomenology}

%%%%%%%%%%%%%%%%%%%%%%%%%%%%%%%%%%%%%%%%%%%%%%%%%%
\begin{figure}[!hbt]
	\centering%
	\includegraphics[width=8cm]{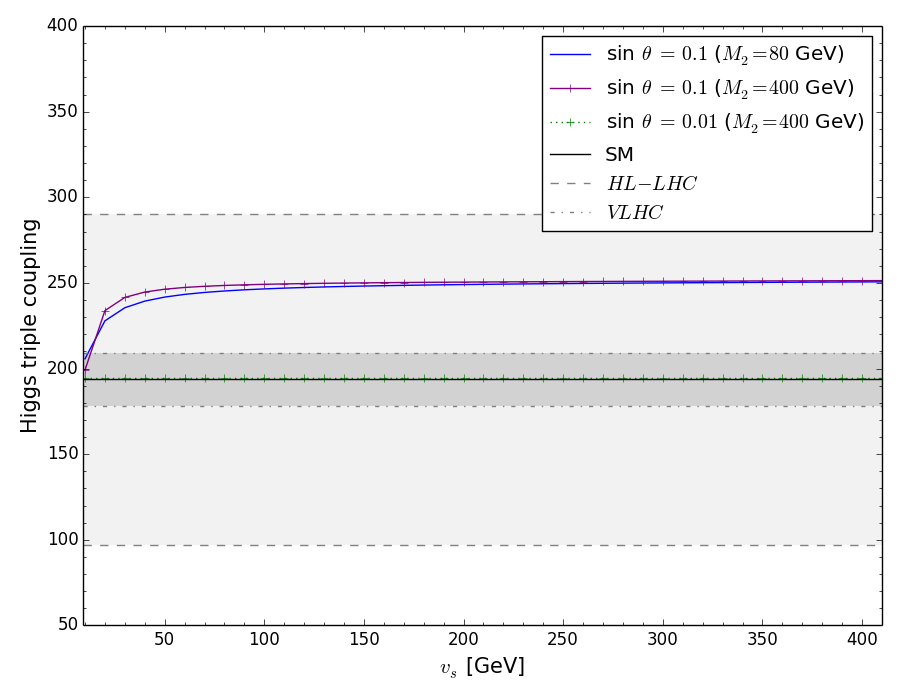}   
	\caption{Higgs triple coupling $c_{111}$ as a function of $v_s$ for different values of $\sin\theta$ and $M_2$.
	Dashed and Dot-dashed lines represent the expected experimental precisions on $c_{111}$ 
    at future hadron colliders, HL-LHC and VLHC, respectively \cite{LHCHiggs14}.   } 
	\label{Branchingf}
\end{figure}
%%%%%%%%%%%%%%%%%%%%%%%%%%%%%%%%%%%%%%%%%%%%%%%%%%

Due to the Higgs portal terms in Eq. \ref{eq:VS_potential}, 
electroweak interaction of the Higgs boson can be significantly modified,
and it is possible that $h_{1,2}$ decay into one another depending on their masses.
If $M_2 \leq M_1/2$, it is kinematically allowed that $h_1$ decays into a pair of $h_2$ 
so that the total decay width of $h_1$ increases.
However, in this case, we found that the total decay width of $h_1$ exceeds too much of
the currently known value of the Higgs decay width $\Gamma_\textrm{SM}$ = 4.07 MeV \cite{LHCHiggs13},
for any value of $\sin\theta$ unless $\lambda_2$ is extremely small.
Therefore, we only consider a case of heavy scalar boson $h_2$ with mass $M_2 > M_1/2$. 

As well as the Higgs portal terms create new interactions between scalar bosons,
they modify the Higgs self-couplings sizably. 
For instance, the Higgs triple coupling $c_{111}$ for $h_1^3$ interaction is given by the above model parameters as
\beq
c_{111} = 6\lambda_0 v_h \cos^3\theta + 3(\lambda_1 + 2\lambda_2 v_s)\cos^2\theta \sin \theta 
+ 6\lambda_2 v_h\cos \theta \sin^2\theta +(\lambda_3 + \lambda_4 v_s)\sin^3\theta ,
\eeq
and can be probed in double Higgs production at hadron or lepton colliders.
One can see from the above equation that the coupling $c_{111}$ reduces to the well-known SM value when $\theta = 0$.
The remaining triple-scalar couplings for $h_1^2 h_2$, $h_1 h_2^2$, and $h_2^3$ interactions 
can be found in Ref. \cite{Kim16}.  
In Fig. \ref{Branchingf}, we illustrate the Higgs triple coupling $c_{111}$ as a function of the scalar VEV $v_s$
for $\sin\theta$ = 0.1, 0.01, 0.001 and for $M_2$ = 80, 400 GeV.
The deviation of experimental value of $c_{111}$ from the SM expectation for $\sin\theta$ = 0.1 lies  
within the expected precision of VLHC experiment, but not within HL-LHC precision. 

%%%%%%%%%%%%%%%%%%%%%%%%%%%%%%%%%%%%%%%%%%%%%%%%%%
\begin{figure}[!hbt]
	\centering%
	\includegraphics[width=8cm]{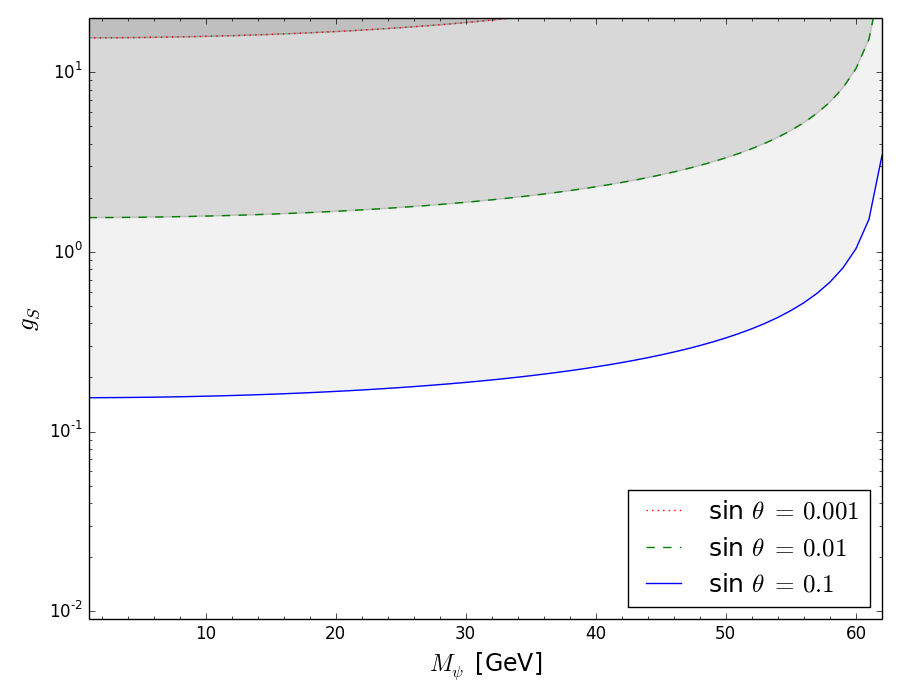}
	\caption{Shaded regions are excluded by the 90\% CL limit of BR($h_1 \to$ inv.) $< 0.20$.
		The upper bounds are shown as different lines for three values of 
		$\sin\theta$ = 0.1(solid blue), 0.01(dashed green), 0.001(dotted red). } 
	\label{fig:invhiggs}
\end{figure}
%%%%%%%%%%%%%%%%%%%%%%%%%%%%%%%%%%%%%%%%%%%%%%%%%%

If $M_\psi < M_1/2$, still, the SM Higgs $h_1$ can decay invisibly into a pair of DM through mixing with decay width,
\beq
\Gamma_\textrm{inv.} = \Gamma\left(h_1 \to \psi\bar{\psi}\right)
= \frac{g_S^2M_1\sin^2\theta}{8\pi}\left(1 - \frac{4M_\psi^2}{M_1^2}\right)^{3/2} ,
\eeq
and the corresponding branching fraction of the invisible Higgs decays is given by the relation
BR($h_1 \to$ inv.) = $\Gamma_\textrm{inv.}/(\Gamma_\textrm{SM} + \Gamma_\textrm{inv.})$,
where $\Gamma_\textrm{SM}$ = 4.07 MeV \cite{LHCHiggs13}.
The most recent upper limit on the Higgs invisible decay has been set 
by the CMS Collaboration combined with the run II data set with the luminosity of 2.3 fb$^{-1}$ 
at center-of-mass energy of 13 TeV \cite{CMS16, Higgsinv}. 
We applied the combined 90\% CL limit of BR($h_1 \to$ inv.) $< 0.20$ to our model, 
and depicted the result for $\sin\theta = 0.1, 0.01, 0.001$ in Fig. \ref{fig:invhiggs}. 
The bounds obtained from the invisible Higgs decays shown in Fig. \ref{fig:invhiggs} are not stronger 
than those from the DM relic density observation, 
and we will discuss it in the next section.
Besides the invisible Higgs decays, due to the vectorial nature and degeneracy of the DM singlet fermion, 
the constraints coming from oblique S,T,U parameters are negligible for $\sin\theta < 0.3$ (within the LEP II constraint)
\cite{Chakraborty13, Baek12}.

%%%%%%%%%%%%%%%%%%%%%%%%%%%%%%%%%%%%%%%%%%%%%%%%%%%%%%%%%%%%%%%%%%%%%%%%
\section{Dark Matter Phenomenology}

The combined results from the recent CMB data by the Planck experiment plus WMAP temperature polarization data gives \cite{PDG14}
\beq
\Omega_{CDM}h^2 = 0.1198 \pm 0.0026.
\label{eq:relic_obserb}
\eeq
This relic density observation will exclude some regions in the model parameter space.
The relic density analysis in this chapter includes all possible channels of
$\overline{\psi}\psi$ pair annihilation into the SM particles.  	
In this work, we implement the model described in section 2 in the FeynRules package \cite{feynrules} 
in order to make use of the MadGraph5 platform \cite{madgraph5}.
Using the numerical package MadDM \cite{maddm2} 
which utilizes the MadGraph5 for computing the relevant annihilation cross sections, 
we obtain the DM relic density 
and the spin-independent DM-nucleon scattering cross sections.

%%%%%%%%%%%%%%%%%%%%%%%%%%%%%%%%%%%%%%%%%%%%%%%%%%
\begin{figure}[!hbt]
	\centering%
	\subfigure[ ]{\label{omegahr1a} %
		\includegraphics[width=7.6cm]{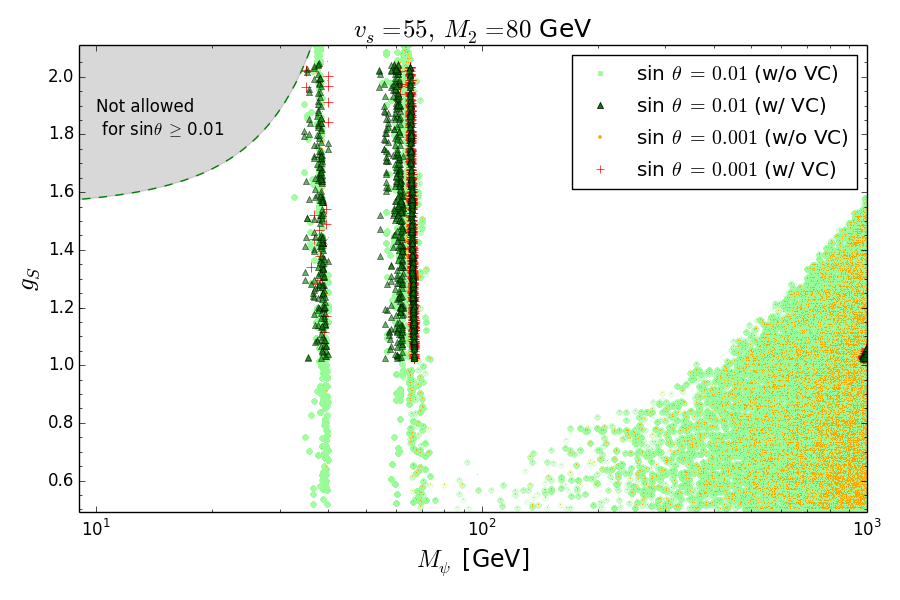}} \
	\subfigure[ ]{\label{omegahr1b} %
		\includegraphics[width=7.6cm]{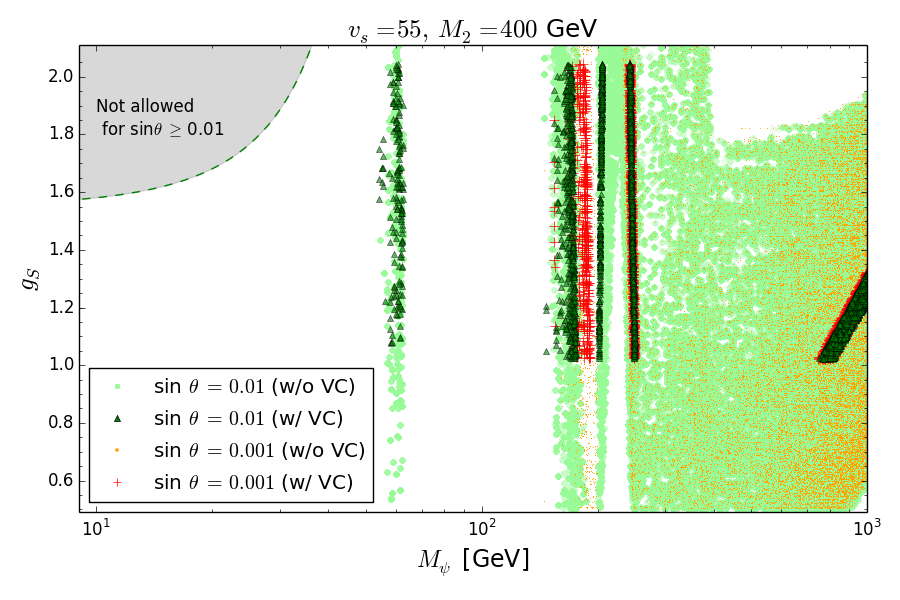}} \\   
	\subfigure[ ]{\label{omegahr2a} %
		\includegraphics[width=7.6cm]{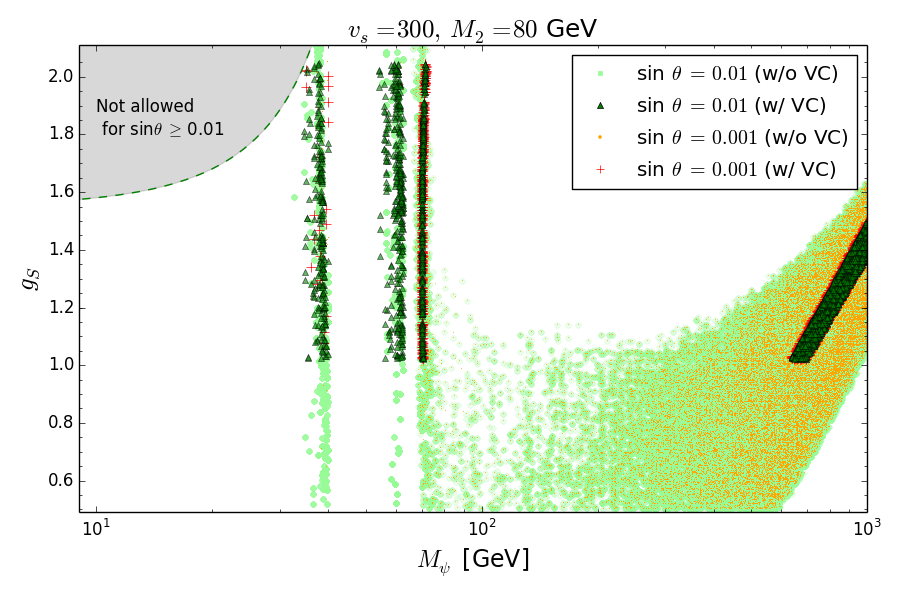}} \
	\subfigure[ ]{\label{omegahr2b} %
		\includegraphics[width=7.6cm]{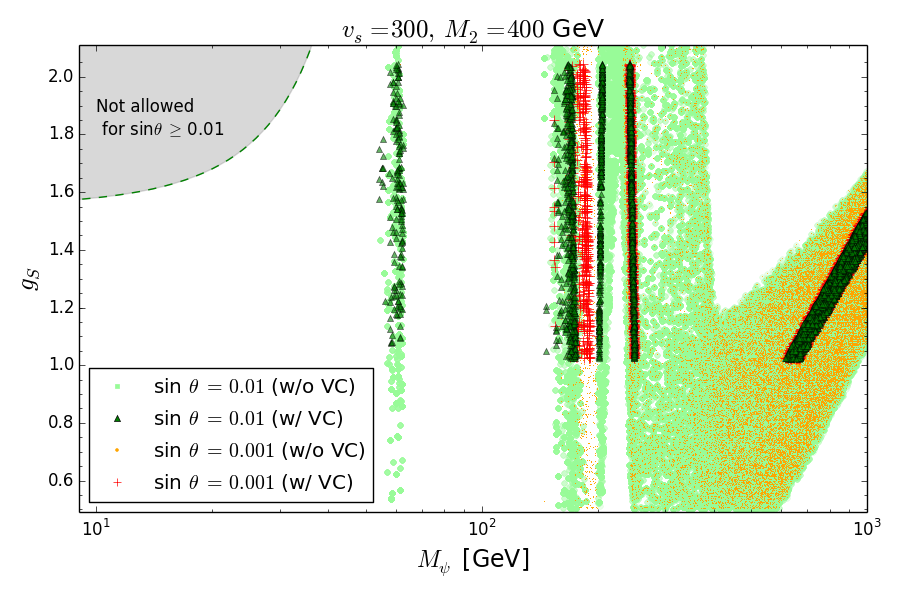}}     
	\caption{Allowed parameter sets of ($M_\psi, g_S$) by relic density observations at $3\sigma$ level
		for $\sin\theta = 0.01, 0.001$ with and without applying the VC.  
		The shaded area above the dashed curve indicates the exclusion limit 
		for $\sin\theta \geq 0.01$ obtained in Fig. \ref{fig:invhiggs}. } 
	\label{fig:relicdensity}
\end{figure}
%%%%%%%%%%%%%%%%%%%%%%%%%%%%%%%%%%%%%%%%%%%%%%%%%%

Instead of performing a complete analysis by varying all relevant independent NP parameters 
($\sin\theta, M_2, M_\psi, g_S, v_s, \lambda_{2,4}$) in this model, 
we choose the benchmark points for the scalar mixing angle $\theta$, 
mass $M_2$, and VEV $v_s$ as follows:
$\sin\theta = 0.1, 0.01, 0.001$, $M_2 = 80, 400$ GeV, and $v_s = 55, 300$ GeV
as similarly considered in Ref. \cite{PMS}
%where implications of the scalar mediator for the EWPT were taken into account
in order to adopt the conditions 
on the NP parameters that lead to a strong first order
phase transition as needed to produce the observed baryon asymmetry of the universe. 
If the VC is applied, $\lambda_{2}$ is fixed to be about 4.17 as obtained in Eq. (\ref{eq:VC_Higgs}),
and $\lambda_{4}$ is determined by the choice of the DM coupling $g_S$ as in Eq. (\ref{eq:VC_scalar})
which is constrained by the unitarity conditions as in Eq. (\ref{eq:VC_coupling}).
If only the unitarity constraints are considered without the VC,
the value of the coupling $g_S$ is arbitrary.
In order to clearly see how the VC constraints the allowed parameter space, 
we also scan the whole perturbative parameter region given in Eq. (\ref{eq:unitarity}) 
and compare the results obtained with and without the VC applied.

For illustration of the relic density constraints on the singlet fermionic DM interaction,
we first plot the allowed region of the DM mass $M_\psi$ and coupling $g_S$ 
constrained by the current relic density observations at $3\sigma$ level
for $\sin\theta = 0.01, 0.001$ in Fig. \ref{fig:relicdensity}, 
where four different sets of new parameters ($v_s$ and $M_2$) are chosen 
in the range of $0.5 < g_S < 2.11$. 
Without the VC constraints, of course, $g_S$ can possibly have a much smaller or larger value,
but we don't consider such a case which digresses from the topic of this paper.     
In Fig. \ref{fig:relicdensity}, one can clearly see that 
the allowed DM parameter space is quite large for arbitrary values of $\lambda_{2,4}$
but much restricted to a small area once we apply the VC and the unitarity condition 
obtained in Eqs. (\ref{eq:VC_Higgs}-\ref{eq:VC_coupling}).
Under the VC, the allowed parameter sets are located near the resonance region of
$M_1$ and $M_2$ or in the mass region of $M_\psi$ heavier than around 700 GeV.    
For different values of $\sin\theta$, 
the allowed parameter regions by the relic density observation are largely overlapped.
For $\sin\theta = 0.01, 0.001$, 
the exclusion limits from the Higgs invisible decay shown in Fig. \ref{fig:invhiggs}
do not give further constraints.
In the case of $\sin\theta = 0.1$, however, 
the allowed region by the relic density observation in the range of $0.5 < g_S < 2.11$ is excluded by 
the spin-independent DM-nucleon scattering result of recent LUX experiment
as we will see later in this chapter, 
so we do not show the result here. 

%%%%%%%%%%%%%%%%%%%%%%%%%%%%%%%%%%%%%%%%%%%%%%%%%%
\begin{figure}[!hbt]
	\centering%
	\includegraphics[width=8cm]{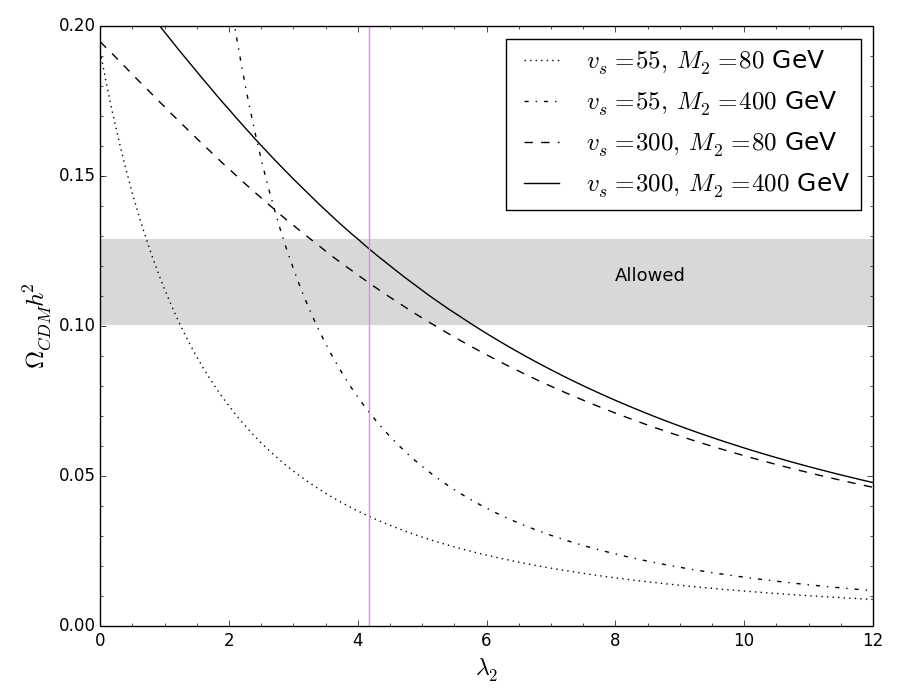}
	\caption{Relic density $\Omega_{CDM}h^2$ as a function of $\lambda_2$ 
		for $\lambda_4 = 3.18$,  $g_S = 1.2$, $M_\psi =800$ GeV, 
		and four different sets of $v_s$ and $M_2$.
		The vertical(violet) line indicates the VC constrained value obtained in Eq. (\ref{eq:VC_Higgs}), 
		and the shaded region is allowed by the recent observation given in Eq. (\ref{eq:relic_obserb}).
		Note that different $\sin\theta$ ($\lesssim 0.1$) doesn't change the plots noticeably in this parameter set 
		because $\sin\theta$ dependence is almost negligible in the heavy $M_\psi$ region. 
	}
	\label{fig:lambda2}
\end{figure}
%%%%%%%%%%%%%%%%%%%%%%%%%%%%%%%%%%%%%%%%%%%%%%%%%%

In the small mass region of $M_\psi$ where DM pair annihilations into $h_1 h_2$ and $h_2 h_2$ 
are not kinematically allowed in the early universe, 
the annihilation cross section is usually quite small (for the small $\sin\theta$ and $g_S$ values) 
so that only near Higgs resonance regions annihilation cross sections can be sizable 
enough to have the correct DM relic density.
But if the DM pair annihilations into $h_1 h_2$ and $h_2 h_2$ are open, 
then the annihilation channels can be dominant  
since the DM pair annihilations into $h_1 h_2$ and $h_2 h_2$ through s-channel $h_2$ mediation 
are not suppressed even with the small $\sin\theta$.
In Fig. \ref{omegahr1b} and \ref{omegahr2b}, 
one can clearly see that the $h_1 h_2$ channel is open for $M_\psi \gtrsim 260$ GeV (average of $M_1$ and $M_2$)
and the $h_2 h_2$ channel also open for $M_\psi \gtrsim 400$ GeV.  
For the annihilation into $h_2 h_2$, the annihilation diagram through t-channel $\psi$ mediation 
is also not suppressed with the small $\sin\theta$. 
The Higgs triple coupling of $h_1 h_2^2$ interaction is given as $c_{122}=2\lambda_2 v_h$ 
in the $\sin\theta \rightarrow 0$ limit. 
Therefore the annihilation cross section (in turn, the relic density) dependence on $\lambda_{2,4}$ 
becomes more important in the heavy $M_\psi$ region.
In Fig. \ref{fig:lambda2}, we show the relic density as a function of $\lambda_2$ for 
$\lambda_4 = 3.18$,  $g_S = 1.2$, $M_\psi =800$ GeV, and four different sets of $v_s$ and $M_2$.
As $\lambda_2$ grows, the Higgs triple coupling $c_{122}$ gets larger 
so that the annihilation cross section gets larger and in turn the relic density gets smaller.
Therefore, if $\lambda_2$ is a free parameter we can adjust the $\lambda_2$ value to have a correct DM relic density. 
However, with the VC imposed, we do not have such freedom (here $\lambda_2$ is fixed as 4.17) 
and therefore only limited parameter sets are survived.

%%%%%%%%%%%%%%%%%%%%%%%%%%%%%%%%%%%%%%%%%%%%%%%%%%
\begin{figure}[!hbt]
	\centering%
	\subfigure[ ]{\label{sigmaSIr1a} %
		\includegraphics[width=7.6cm]{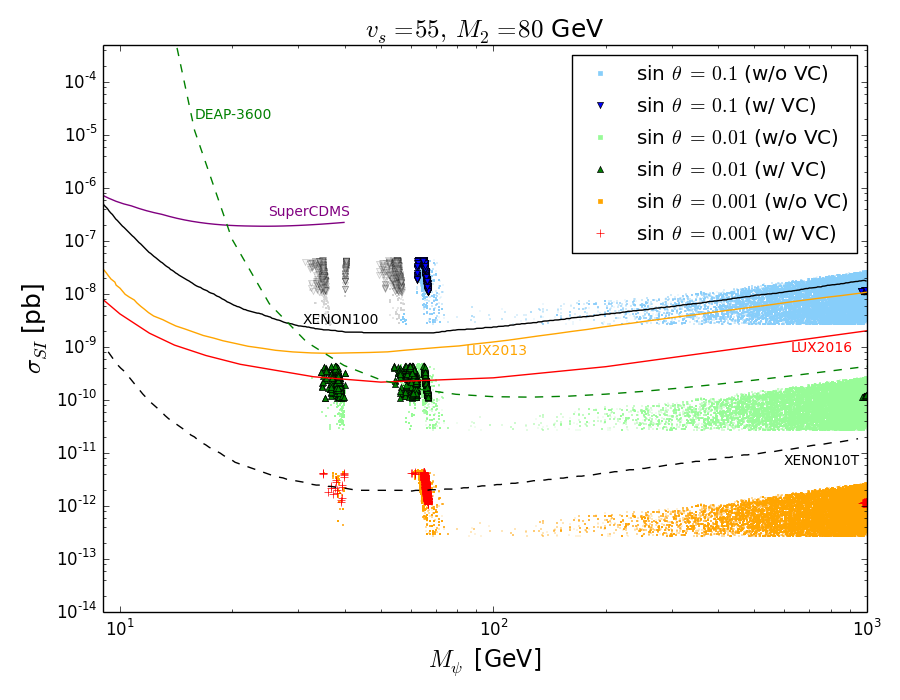}}  \
	\subfigure[ ]{\label{sigmaSIr1b} %
		\includegraphics[width=7.6cm]{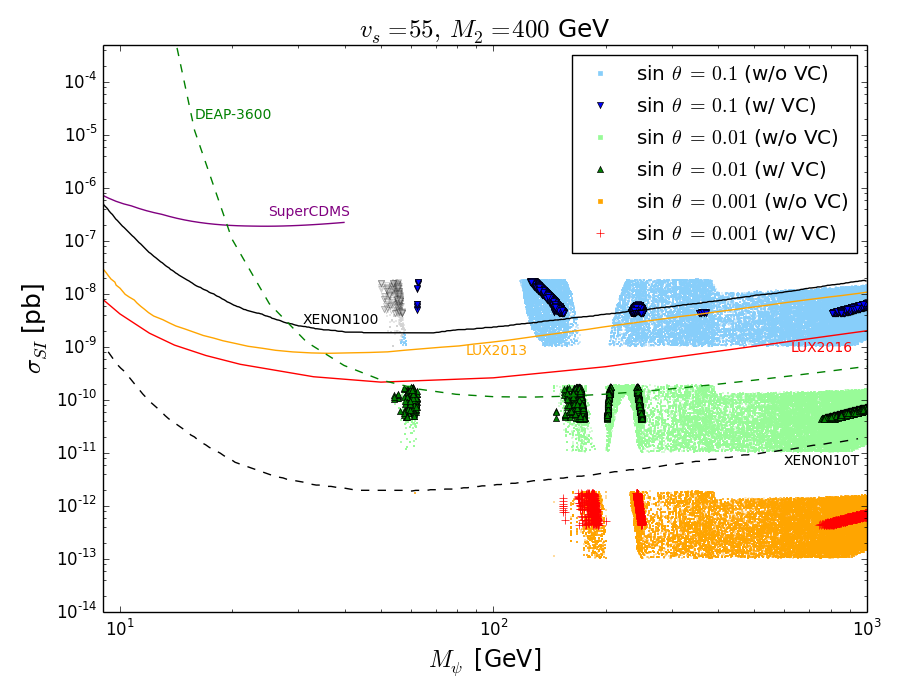}}   \\
	\subfigure[ ]{\label{sigmaSIr2a} %
		\includegraphics[width=7.6cm]{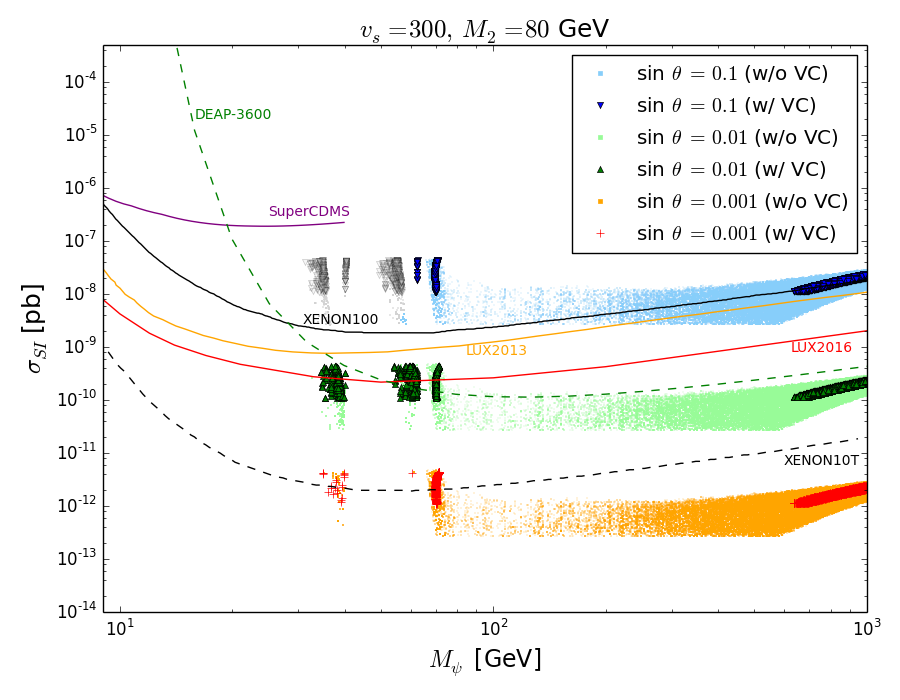}}  \
	\subfigure[ ]{\label{sigmaSIr2b} %
		\includegraphics[width=7.6cm]{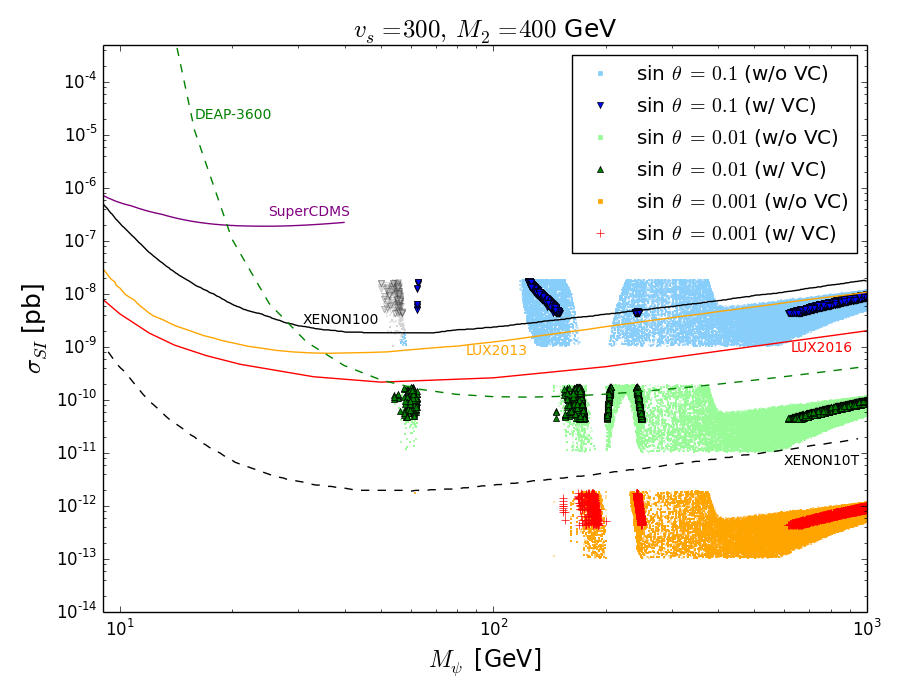}}   
	\caption{Spin-independent DM-nucleon scattering results allowed by relic density observations 
		for $0.5 < g_S < 2.11$ with and without applying the VC.  
		Also shown are observed limits from LUX 2013, 2016, XENON100, SuperCDMS,  
		and expected limits from DEAP-3600 and XENON10T.
	    For $\sin\theta = 0.1$, 
	    some of allowed parameter sets by the relic density observation 
	    in low DM mass (grayed) region  
	    are excluded by the invisible Higgs decay bound given in Fig. \ref{fig:invhiggs}.} 
	\label{fig:sigmaSI}
\end{figure}
%%%%%%%%%%%%%%%%%%%%%%%%%%%%%%%%%%%%%%%%%%%%%%%%%%

In Fig. \ref{fig:sigmaSI}, 
we plot the spin-independent DM-nucleon scattering cross section 
by varying the DM mass $M_\psi$ with parameter sets allowed by the relic density observation,
and compare the results with the observed upper limits obtained at 90$\%$ level 
from LUX 2013, 2016, XENON100, SuperCDMS, 
and with the expected limits from DEAP-3600 and XENON10T. 
One can see from the figure that $\sin\theta \gtrsim 0.1$ case obtained in the range of $0.5 < g_S < 2.11$ 
is not favored in this model due to the recent LUX bound \cite{LUX16},
and most of allowed regions for $\sin\theta = 0.01, 0.001$ are not excluded.
Also, $\sin\theta = 0.01$ scenario can be tested sooner or later by ongoing experiments such as XENON10T. 

If one chooses the second scenario ($\lambda_2(v) = 2.0$) instead of the first scenario 
($\lambda_2(v) = 4.17$),
the allowed regions for $M_\psi > 600$ GeV with the VC applied in Fig. \ref{fig:relicdensity} and \ref{fig:sigmaSI} 
will be shifted a bit but not much, and still be inside of the allowed regions 
obtained generally without the VC. 
Similarly, even if one introduces additional NP in the hidden sector at a TeV scale and apply the VC at the NP scale,
our numerical results can still be applicable and provide meaningful guidelines as discussed in Sec. III.

\section{Concluding Remarks}

In this paper, we studied the renormalizable Higgs portal model of a fermionic dark matter 
with a gauge singlet Dirac fermion and a real single scalar 
which ensure the cancellation of one-loop quadratic divergence 
to the Higgs and the additional singlet scalar masses. 
We showed that our DM model can satisfy the VCs for the quadratic divergences 
to the masses of the SM-like Higgs boson and the singlet scalar simultaneously.
We constrained the Higgs-scalar coupling $\lambda_2$ 
and the scalar self-interaction coupling $\lambda_4$ using the VC.
We further constraint those couplings using the perturbative unitarity conditions,
and obtained the bound on the DM coupling $g_S$ as well. 

We showed the allowed region of NP parameters, the DM mass $M_\psi$ and coupling $g_S$, 
satisfying the recent measurement of relic abundance 
with and without the VC constraints in Fig. \ref{fig:relicdensity}.
We also showed the spin-independent DM-nucleon scattering cross section of the singlet fermionic DM 
by varying the DM mass $M_\psi$ with parameter sets allowed by the relic density observation, 
and compare the results with the observed upper limits from various experiments in Fig. \ref{fig:sigmaSI}.
In the figures, one can clearly see that
the allowed parameter space constrained by the VC are located near the resonance region 
of $M_1$ and $M_2$ or in the mass region of $M_\psi$ heavier than about 700 GeV.
We performed the numerical analysis for three different values of the Higgs-scalar mixing angle
$\sin\theta = 0.1, 0.01, 0.001$, and the $\sin\theta = 0.1$ case is disfavored 
in this model due to the recent LUX bound
even in the case without the VC applied.

Our renormalizable DM model with a singlet Dirac fermion and a real single scalar is a minimal DM model 
which has allowed NP parameter space 
where the fine-tuning (or the hierarchy) problem of the Higgs mass is much relaxed,
and this model can be expanded by introducing complex scalars and/or DM fermion multiplets.
The obtained allowed parameter sets under the VC can be used as benchmark points
to test proper DM model candidates as future experimental progress can further improve the bounds.

\acknowledgments
This work was supported by Basic Science Research Program
through the National Research Foundation of Korea (NRF)
funded by the Ministry of Science, ICT and Future Planning
(Grant No. NRF-2015R1A2A2A01004532 (K.Y. Lee) and NRF-2017R1E1A1A01074699 (S.-h. Nam)), 
and funded by the Ministry of Education (Grant No. NRF-2016R1A6A3A11932830 (S.-h. Nam)).

%%%%%%%%%%%%%%%%%%%%%%%%%%%%%%%%%%%%%%%%%%%%%%%%%%
\def\npb#1#2#3 {Nucl. Phys. B {\bf#1}, #2 (#3)}
\def\plb#1#2#3 {Phys. Lett. B {\bf#1}, #2 (#3)}
\def\prd#1#2#3 {Phys. Rev. D {\bf#1}, #2 (#3)}
\def\jhep#1#2#3 {J. High Energy Phys. {\bf#1}, #2 (#3)}
\def\jpg#1#2#3 {J. Phys. G {\bf#1}, #2 (#3)}
\def\epj#1#2#3 {Eur. Phys. J. C {\bf#1}, #2 (#3)}
\def\arnps#1#2#3 {Ann. Rev. Nucl. Part. Sci. {\bf#1}, #2 (#3)}
\def\ibid#1#2#3 {{\it ibid.} {\bf#1}, #2 (#3)}
\def\none#1#2#3 {{\bf#1}, #2 (#3)}
\def\mpla#1#2#3 {Mod. Phys. Lett. A {\bf#1}, #2 (#3)}
\def\pr#1#2#3 {Phys. Rep. {\bf#1}, #2 (#3)}
\def\prl#1#2#3 {Phys. Rev. Lett. {\bf#1}, #2 (#3)}
\def\ptp#1#2#3 {Prog. Theor. Phys. {\bf#1}, #2 (#3)}
\def\rmp#1#2#3 {Rev. Mod. Phys. {\bf#1}, #2 (#3)}
\def\zpc#1#2#3 {Z. Phys. C {\bf#1}, #2 (#3)}
\def\cpc#1#2#3 {Chin. Phys. C {\bf#1}, #2 (#3)}
\def\jcap#1#2#3 {JCAP {\bf#1}, #2 (#3)}
\def\compc#1#2#3 {Comput. Phys. Commun. {\bf#1}, #2 (#3)}
\def\pdu#1#2#3 {Phys. Dark Univ. {\bf#1}, #2 (#3)}
%%%%%%%%%%%%%%%%%%%%%%%%%%%%%%%%%%%%%%%%%%%%%%%%%%

\end{document}